\documentclass{SCGE}
\usepackage[colorlinks,linkcolor=blue,citecolor=blue,urlcolor=blue]{hyperref}
\usepackage{graphicx} 
\usepackage{epstopdf}
\usepackage[T1]{fontenc}
\usepackage{ulem}

\let\citedash\relax
\makeatletter \providecommand{\citedash}{\hbox{-}\penalty\@m}
\makeatother

\begin{document}
\begin{picture}(0,0){\rm
\put(0,-20){\makebox[160truemm][l]{\bf {\sanhao\raisebox{2pt}{.}}
Article  {\sanhao\raisebox{1.5pt}{.}}}}}
\put(0,-34){\jiuwuhao {\textcolor[rgb]{0.5,0.5,0.5}{\sf 
}}}
\end{picture}

\def\bm{\boldsymbol}

\def\dl{\displaystyle}
\def\du{\end{document}}
\def\d{{\rm d}}
\def\e{{\rm e}}
\def\i{{\rm i}}

\Year{2019} %
\Month{May} %
\Vol{62} 
\No{5} 
\BeginPage{1} 
\AuthorMark{{\rm H. Wang}, et al.}  
\DOI{} 
\ArtNo{959507}

\title[Pulsar Candidate Selection Using Ensemble Networks for FAST Drift-Scan Survey]{Pulsar Candidate Selection Using Ensemble Networks for FAST Drift-Scan Survey}

\author[1,2,3]{Hongfeng Wang}{hfwang@mail.bnu.edu.cn}
\author[2]{Weiwei Zhu}{zhuww@nao.cas.cn}
\author[4]{Ping Guo}{pguo@ieee.org}
\author[2,5]{Di Li}{}
\author[1]{Sibo Feng}{}
\author[1]{Qian Yin}{}
\author[5,2]{\\Chenchen Miao}{}
\author[7,2]{Zhenzhao Tao}{}
\author[2]{Zhichen Pan}{}
\author[2]{Pei Wang}{}
\author[1]{Xin Zheng}{}
\author[4]{Xiaodan Deng}{}
\author[6]{\\Zhijie Liu}{}
\author[6]{Xiaoyao Xie}{}
\author[6]{Xuhong Yu}{}
\author[6]{Shanping You}{}
\author[6]{Hui Zhang}{}
\author[]{FAST Collaboration}{}

\address[{\rm1}]{Image Processing and Pattern Recognition Laboratory, College of Information Science and Technology, Beijing Normal University, Beijing 100875, China;}
\address[{\rm2}]{CAS Key Laboratory of FAST, Chinese Academy of Science, Beijing 100101, China;}
\address[{\rm3}]{School of Information Management, Dezhou University, Dezhou 253023, China;}
\address[{\rm4}]{Image Processing and Pattern Recognition Laboratory, School of Systems Science, Beijing Normal University, Beijing 100875, China;}
\address[{\rm5}]{University of Chinese Academy of Sciences, Beijing 100049, China;}
\address[{\rm6}]{Key Laboratory of Information and Computing Science Guizhou Province, Guizhou Normal University, Guiyang 550001, China;}
\address[{\rm7}]{School of Physics and Electronic Science, Guizhou Normal University, Guiyang 550001, China;}

\maketitle \vspace{-3.5mm}{\footnotesize\begin{center} 
\end{center}}\vspace*{-5mm}

\begin{center}
\rule{16.5cm}{0.4pt}
\parbox{16.5cm}
{\begin{abstract}
The Commensal Radio Astronomy Five-hundred-meter Aperture Spherical radio Telescope (FAST) Survey (CRAFTS) utilizes the novel drift-scan commensal survey mode of FAST and can generate billions of pulsar candidate signals. The human experts are not likely to thoroughly examine these signals, and various machine sorting methods are used to aid the classification of the FAST candidates. In this study, we propose a new ensemble classification system for pulsar candidates. This system denotes the further development of the pulsar image-based classification system (PICS), which was used in the Arecibo Telescope pulsar survey, and has been retrained and customized for the FAST drift-scan survey. In this study, we designed a residual network model comprising 15 layers to replace the convolutional neural networks (CNNs) in PICS. The results of this study demonstrate that the new model can sort $>$96\% of real pulsars to belong the top 1\% of all candidates and classify $>1.6$ million candidates per day using a dual--GPU and 24--core computer. This increased speed and efficiency can help to facilitate real-time or quasi-real-time processing of the pulsar-search data stream obtained from CRAFTS. In addition, we have published the labeled FAST data used in this study online, which can aid in the development of new deep learning techniques for performing pulsar searches.
\end{abstract}}
\end{center}\vspace*{-0.6cm}

\begin{center}
\parbox{16.5cm}
{\bf\jiuhao Pulsars, Neural networks, Data analysis}
\end{center}

\begin{center}
{}
\CITA    
\end{center}

\textwidth=178truemm \textheight=236truemm

\wuhao\vspace*{1.5mm}

\begin{multicols}{2}

\renewcommand{\baselinestretch}{1.08} \baselineskip 12.2pt\parindent=10.8pt

\renewcommand{\thefootnote}

\section{Introduction}\label{sec:1}
Pulsars are rapidly rotating neutron stars that emit radio wave pulses via their strong magnetic fields. They serve as extraordinary astrophysical laboratories for studying both nuclear physics and strong gravitation. Thus far, less than 2,700 pulsars have been discovered throughout the galaxy, and the majority of these pulsars have been discovered through pulsar surveys using radio telescopes. A large number of pulsars have been discovered in modern pulsar surveys, including the
Parkes multi-beam pulsar survey (PMPS; \cite{manchester2001parkes}), the high time resolution universe (HTRU) survey \cite{burke2011high}, the pulsar Arecibo L-band feed array (PALFA) survey \cite{deneva2009arecibo}, the Green Bank Telescope (GBT) drift-scan pulsar survey \cite{Boyles2012The}, the Green Bank North Celestial Cap (GBNCC) survey \cite{GBNCC}, and the low-frequency array (LOFAR) tied-array all-sky survey (LOTAAS; \cite{LOTAAS}). Modern pulsar surveys often produce a large number of potential candidates; however, only a small fraction of these candidates are actual pulsars. For example, the HTRU survey produced several million pulsar candidates, and the GBT drift-scan pulsar observation project produced more than 1.2 million pulsar candidates \cite{Boyles2012The, Lynch2012The}. The Five-hundred-meter Aperture Spherical radio Telescope (FAST) was installed on September 25, 2016, and is currently being commissioned \cite{jyg+19}. The FAST science team used a single--pixel feed and conducted an initial drift-scan pulsar survey, producing more than 10 million candidates before May 2018. Since then, a FAST 19--beam receiver has been installed, generating more than one million pulsar candidates per night. It is infeasible for such a large number of candidates to be individually examined  by several experts; therefore, we propose a deep-learning--based pulsar candidate selection system. This system has been modified based on the pulsar image-based classification system (PICS) and customized for conducting the FAST drift-scan survey. We have designed a ResNet model comprising 15 layers to replace the convolutional neural networks (CNNs) in PICS. The resulting model can run on both the CPU and GPU platforms and surpasses PICS in terms of both effectiveness and runtime.

The Commensal Radio Astronomy FasT Survey (CRAFTS; \cite{CRAFTS}) is a multi-objective drift-scan survey that has been conducted based on the FAST. We intend to use the FAST 19--beam L-band receiver in drift mode to observe the visible sky of the FAST for HI emission and new pulsars. The drift-scan mode survey produces a large number of pulsar candidates because of the large survey time and small integration time ($\sim$12s); additionally, it  generates pulsar candidates that emerge and subsequently disappear as the pulsars drift away. These candidates may differ from the canonical pulsar candidates that are often used for training previous pulsar classification systems. The radio frequency interference (RFI) environment of the FAST also differs from that of other telescopes. Therefore, it is crucial to train a new pulsar classifier for conducting this new large-scale survey. This study utilizes the pulsar and RFI data obtained from a previous work and other telescopes and combines them with the new data obtained from the FAST survey. This serves to produce a novel system that is intend to be more generalized than the PICS, which is its predecessor.

In Section 2, we describe the pipeline of the pulsar candidates and review various types of automatic pulsar candidate selection. We further introduce the ResNet model, which is our method. In Section 3, we introduce the FAST, PALFA, and HTRU pulsar datasets and describe the manner in which we conducted the experiments using the data obtained from the GBNCC and FAST. Our experimental results denote that the proposed pulsar candidate selection method runs efficiently and exhibits high recognition accuracy. Finally, Section 4 presents the discussions and conclusions.

\section{Machine learning method for pulsar candidate selection \label{sec:2}}
In this section, we explain the generation of pulsar candidates, summarize a previous study related to automatic pulsar candidate selection, and discuss the design of our 15-layer ResNet model.

\subsection{Pulsar candidates and selection methods}

In their study, Lyon et al. (2016) \cite{lyon2016fifty} described a general procedure for detecting pulsars. A telescope system collects radio signals over a large number of frequency channels at a high time resolution (usually < 1 $ms$ of the sample time). The RFI--contaminated frequency channels are eliminated, and the signals are further de-dispersion to form the time series for several different dispersion trails. Dispersion refers to the transmission of radio signals having different frequencies through the interstellar medium while encountering various delays. These delays, if not accounted for, can reduce the signal--to--noise ratio (S/N) of the pulsar signal.
For an unknown pulsar, the exact dispersion value is unknown \cite{M2010The}. Therefore, several different dispersion measures (DMs) should be considered. The subsequent step involves searching for the periodic signals in the time series of different DMs using Fourier analysis. Further, a fast Fourier transform (FFT) of the time series data can be obtained, and the results can be examined for the presence of significant power peaks.
Some pulsars have considerably narrow pulses. The FFT of narrow pulses tends to spread to multiple harmonic frequencies, and this problem can be addressed by harmonic summing, which is the subsequent step during pulsar search. A maximum of 32 harmonics are often summed to produce the highest signal peaks. The signals that are detected with sufficiently large Fourier power are considered to be the candidate signals. Majority of these candidates are related to an RFI or known periodicity and can be referred to as birdies. The signals that are related to these birdies or that have RFI-like distributions are discarded during this stage. Finally, a large number of remaining candidates are retained for performing further analysis. Further, the diagnostic plots and summary statistics are calculated for these candidate signals by folding the original data using each signal's DM and period. The diagnostic plots contain a small set of features that include the signal's S/N, DM, period, pulse width, integrated pulse profile, and two-dimensional (2D) plots that display the signals' variance with respect to time, frequency, and DM.

\begin{figure}[H]
\centering
\includegraphics[width=0.9\linewidth]{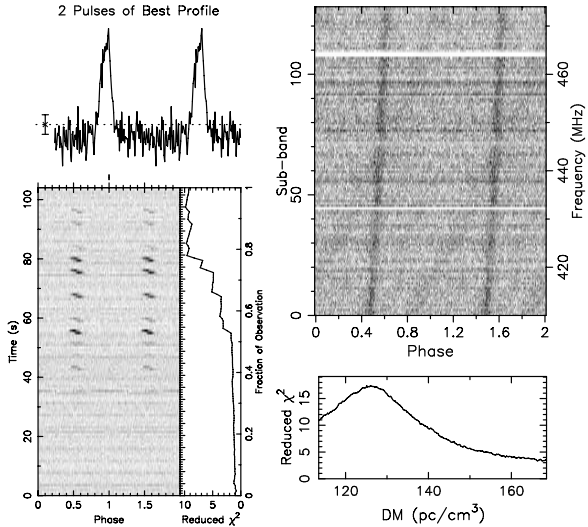}
\caption{Real pulsar candidate obtained from the FAST drift-scan survey. The time domain displays an intermittent signal caused by the pulsar drifting in and out of the beam. The phase versus time plot reveals a significant subpulse drift in the pulsar. }
\label{fig:pulsar}
\end{figure}

\begin{figure}[H]
\centering
\includegraphics[width=0.9\linewidth]{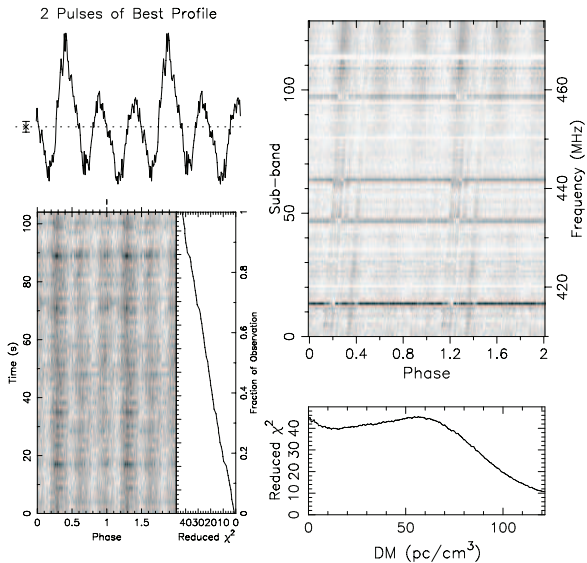}
\caption{The example of a non-pulsar candidate obtained from the FAST data. The frequency versus phase plot contains features that resemble the pulsar signals. However, upon close examination, these features become entirely vertical when the plot is refolded with DM set to zero, demonstrating that the signal is not dispersed. Thus, this candidate is most likely to have originated from the ground interference.}
\label{fig:non_pulsar}
\end{figure}

We used the PulsaR Exploration and Search TOolkit (\textsc{presto}) pipeline software for implementing the aforementioned steps. This software generates candidate diagnostic plots.  Figure~\ref{fig:pulsar} presents an example of a pulsar candidate plot based on the FAST drift-scan survey. 
Further, we can examine the candidates using the following four important subplots: the summed profile plot, the phase versus time plot, the phase versus frequency plot, and the DM curve plot. These plots are the four fundamental features that are considered by the human experts to classify the candidates. The positive (pulsar, Figure~\ref{fig:pulsar}) and negative (non-pulsar, Figure~\ref{fig:non_pulsar}) candidates exhibit different characteristics.

Recently, there has been an increase in the usage of computational intelligence and machine learning technology in various fields of astronomy\cite{wang2018cis}. Several researchers have successfully applied machine learning, which has significantly improved the efficiency of the pulsar searches, to perform pulsar candidate selection. Eatough et al. (2010) \cite{eatough2010selection} extracted 12 features of pulsar candidates and trained a single-hidden-layer artificial neural network, discovering a new pulsar in their PMPS data. Lee et al. (2013) \cite{lee2013peace} proposed a pulsar scoring method that can be referred to as the pulsar evaluation algorithm for candidate extraction (PEACE), which focuses on the quantitative evaluation of six features of a candidate and determines whether this candidate is a pulsar. Bates et al. (2012) \cite{Bates2012The} adopted 22 features and trained a single-hidden-layer artificial neural network.  Morello et al. (2014) \cite{morello2014spinn} proposed straightforward pulsar identification using neural networks (SPINN), whose design contained six artificial features in the input layer, eight neurons in the hidden layers, and one neuron in the output layer.  Lyon et al. (2016)\cite{lyon2016fifty} proposed a pulsar candidate selection method based on a decision tree, which can be referred to as the Gaussian Hellinger very fast decision tree (GH-VFDT) and contains eight features obtained from the diagnostic histograms to describe a pulsar candidate. This model has been successfully applied in LOTAAS. For improving the pulsar recognition accuracy, Guo et al. (2017) \cite{2017arXiv171110339G} proposed DCGAN + L2-SVM to address the class imbalance problem. They used DCGAN to learn features and trained an L2-support vector machine (L2-SVM) model to predict the results. The FRB search method is similar to the pulsar search method; Connor et al. (2018) \cite{Connor(2018)} applied a tree-like deep neural network (DNN) to the FRB searches. Zhang et al. (2018) \cite{2018ApJzhang} were the first to utilize the ResNet architecture for detecting fast radio bursts, and they detected 72 new pulses from FRB 121102 that had not been identified by the S/N-based methods in previous searches \cite{Gajjar2018}.

Artificial neural networks, especially CNNs, have played an important role in earlier pulsar candidate selection. For instance, Zhu et al. (2014)\cite{zhu2014searching} proposed a pulsar image-based classification system. In accordance with the four important features of a pulsar candidate (the summed profile, time versus phase plot, frequency versus phase plot, and DM curve), they developed the design of an ensemble network combined with a single-hidden-layer network, an SVM, and a CNN. Among these, CNNs have demonstrated good classification performance for 2D diagnostic plots. For example,  a CNN is a typical end-to-end model when compared to other methods. In constrast, other methods require hand-crafted features for performing automatic pulsar candidate selection. Therefore, it is necessary for a pulsar expert to design artificial features, which may decrease the efficiency of pulsar candidate selection. However, the objective of this study is to replace the CNN with ResNet in PICS, which is developed using a TensorFlow framework for classifying the efficiency.
\subsection{Residual network design}

A CNN is an excellent feature extractor because it performs end-to-end learning of the raw image data to be classified. In PICS, the CNN model is relatively slow in training and classification because it is not optimized to run on either a CPU or a GPU. In addition, for legitimate pulsar candidates, there are one or more vertical lines in the 2D subplot (time versus phase and frequency versus phase); these features are usually captured in the early layers of the neural network. However, CRAFTS generates a wide range of pulsar candidates, especially those exhibiting sub-pulse drift. Deep networks must be designed to recognize these complex features. Therefore, we have adopted a residual network model \cite{He2016Deep} based on GPU acceleration for performing 2D subplot classification.

\begin{figure}[H]
\centering
\includegraphics[width=0.7\linewidth]{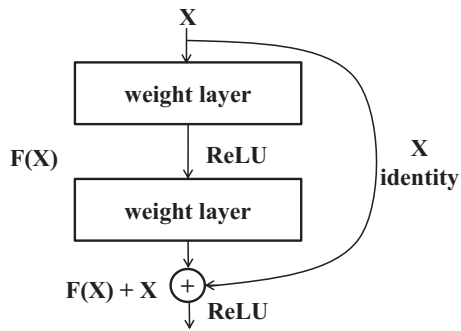}
\caption{Illustration of a residual block. Residual learning: a building block.}
\label{fig:residual_block}
\end{figure}

A residual network can be considered to be a typical CNN. Deep networks tend to have a low back propagation error owing to gradient degradation; this problem arises during the deep network training process. The error term gradually decreases during the process of backward propagation, reducing the learning effectiveness in the top layers of the network. He et al.\cite{He2016Deep} proposed a residual learning framework (Figure ~\ref{fig:residual_block}) to address this degradation problem. The principle of this approach involves an "identity-connection" that connects the blocks of the convolutional layers and computes the shallow and deep features. The identity connections are shortcuts preventing the gradients from vanishing too rapidly. Thus, multiple residual networks can be stacked together to form deep networks without suffering from gradient degradation. In practice, this type of deep residual network can improve the classification accuracy. Hardt et al. (2016) \cite{2016arXiv161104231H}, Du et al. (2018) \cite{2018arXiv181103804D} have also proposed theories for ensuring the effectiveness of the ResNet method.

\begin{figure*}
\centering
\includegraphics[width=\textwidth,scale=0.8]{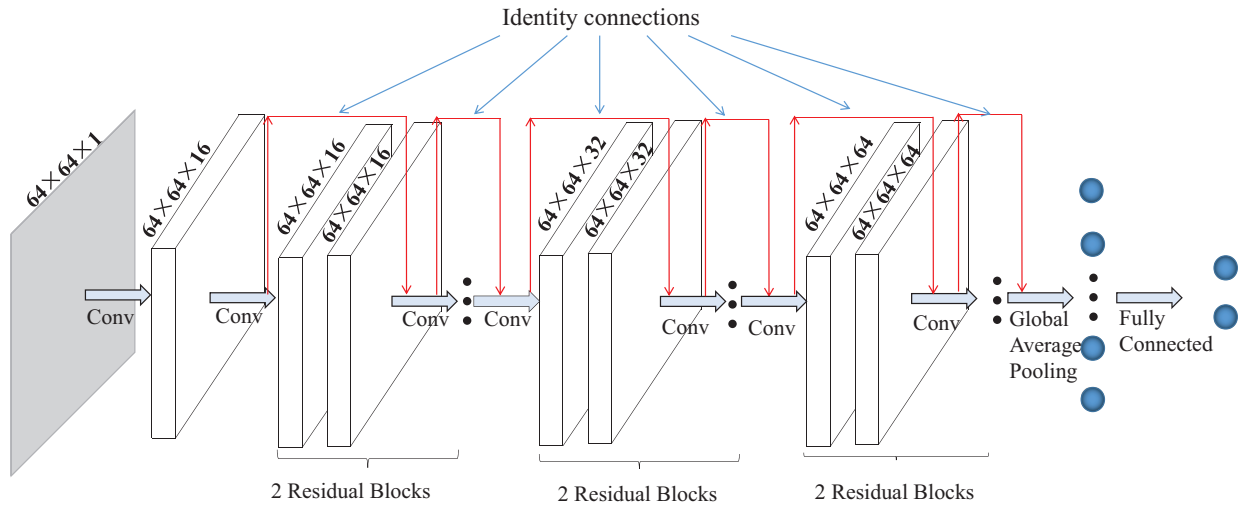}
\caption{Diagram of a 15 layered ResNet model. Conv refers to convolutional operation. The "2 Residual Blocks" component contains two building blocks that possess an identical architecture. The input image is $64 \times 64 \times 1$, and the output sizes are $H \times W \times N$, where H and W denote the height and width of the tensor, and N denotes the number of features.}
\label{fig:ResNet model}
\end{figure*}

We designed the ResNet architecture according to the characteristics of the 2D subplots of pulsar candidates and the experimentation performed using various combinations of hyperparameters. We varied the network depth, feature map numbers, and kernel size and observed the changes in classification accuracy. Further, we determined the network structure and hyperparameters based on multiple factors, including the computation time, classification accuracy, and performance convergence. We conducted an experiment using networks of different depths and discovered that both a shallow 9-layer ResNet and a deep 39-layer ResNet exhibited lower accuracy than that exhibited by a 15-layer ResNet. Thus, our ResNet model comprises 15 layers (Figure~\ref{fig:ResNet model}). 
The network layers include one input convolutional layer, two output layers, and 12 convolutional layers. The 12 convolutional layers are divided into three groups, each group containing two residual blocks and each residual block containing two convolutional layers.
Figure~\ref{fig:residual_block} displays the structure of a residual block; in each residual block, the parameters of the two convolutional layers are identical, and two residual blocks in the same group are observed to be identical to each other. After the residual blocks is the 14~th layer, which is a global average pooling layer that is connected to the 15~th layer, a fully connected softmax output layer. 

While using this model to classify the pulsar candidates, we preprocessed the feature data before feeding them to the neural network. This procedure involved re-scaling the data to the zero mean and unit variance and shifting the peak to a phase of 0.5 for both the pulsar profile and 2D plots; however, we did not shift the peak of the DM curve. 

Figure~\ref{fig:ResNet model} presents the design of the ResNet model used for classifying the 2D images. We extracted four main features plots of a candidate from the pfd files; these plots included one-dimensional (1D) data arrays (summed profile and DM curve) and 2D data arrays (time versus phase and frequency versus phase subplots). Further, the original size of the feature data varied among various candidates. However, because our machine learning models functioned only using input data of identical size, we downsampled or interpolated the data to a uniform size before feeding them to our model. For a 1D subplot, we applied an SVM model in which the size of the input array was $64 \times 1$.  In contrast, for the 2D subplot, we used the ResNet model in which the size of the input array was $64 \times 64$.

\begin{figure*}
\centering
\includegraphics[width=\textwidth,scale=0.5]{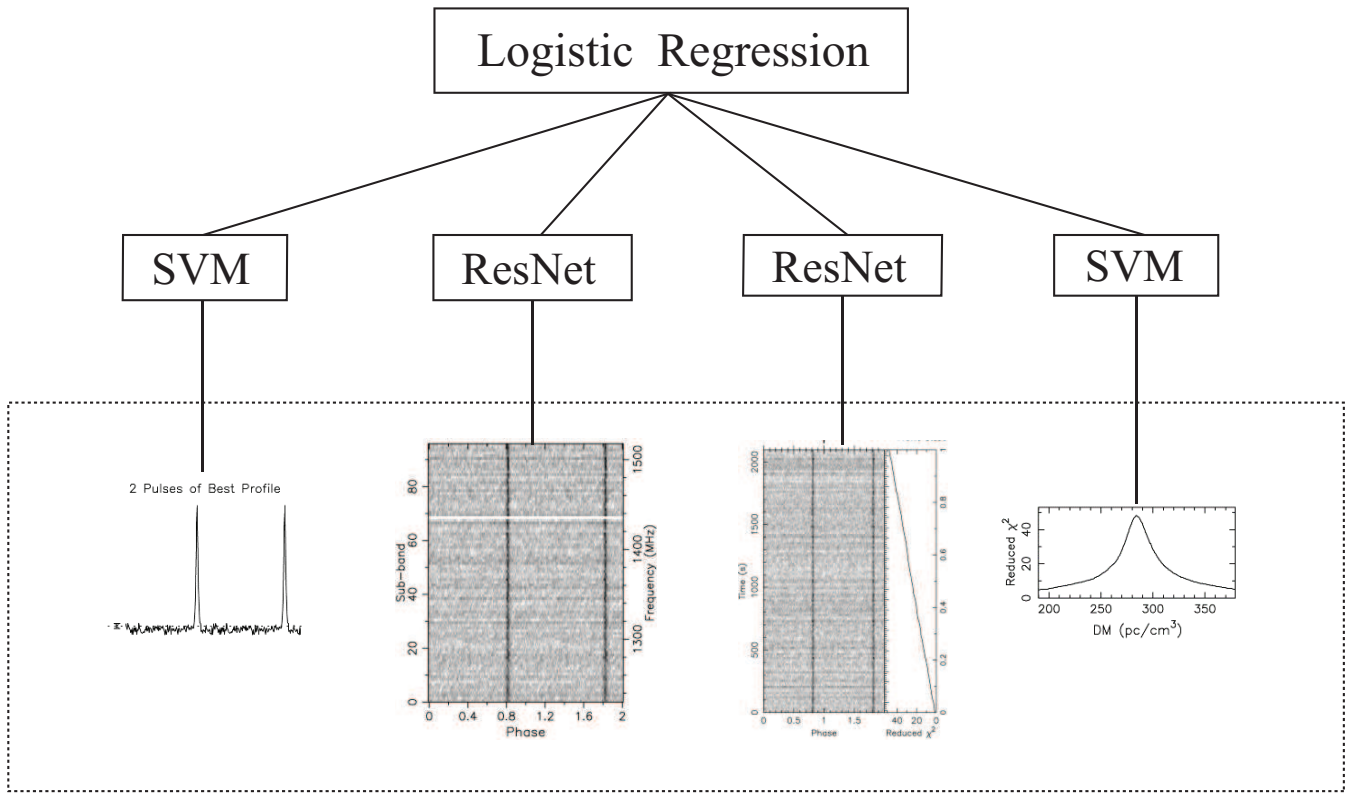}
\caption{Diagram of the PICS--ResNet model. The first layer classifies the individual features (the pulse profile, time versus phase plot, frequency versus phase plot, and DM curve), whereas the second layer classifies the candidates based on the results of the first layer. The SVM components represent the support vector machine model, while the ResNet components represent the residual network model. See Figure~\ref{fig:ResNet model} for a schematic of the ResNet model.}
\label{fig:our method}
\end{figure*}

\section{Results}\label{sec:3}

In this section, we present the training and validation of PICS--ResNet, which is our new model and compare it with its predecessor, which is a generic PICS model. The generic PICS model used in this study exhibits the same architecture as that exhibited by the model described by Zhu et al. (2014) \cite{zhu2014searching} but with several differences. The original PICS model used a Bayesian prior to reject RFIs based on the frequency population of all the candidates (Lee et al. 2013 \cite{lee2013peace}; Zhu et al. 2014 \cite{zhu2014searching}). The PICS model in this study is trained with a larger amount of data than that used to train the original model; however, it does not reject RFIs based on their periods. Further, the results of the following three experiments designed to test robustness of the models are presented: 1. training both the models using an old dataset and testing them using the GBNCC dataset; 2. training both the models using a new dataset, including the FAST data, and testing them using the GBNCC dataset; 3. testing the models trained with the FAST data using a small set of reserved FAST test data. The results demonstrate that the PICS--ResNet model converges better in training and outperforms the PICS model in testing. Both the models display improvement after being trained with new data obtained from the FAST, exhibiting a recall rate of more than 90\% . 

\subsection{Datasets and evaluation metrics}

The pulsar candidate datasets that were used included the PALFA, HTRU, GBNCC, and FAST datasets (Table \ref{tab:table0}). Some of the datasets, such as the PALFA and GBNCC datasets, were used and presented in a previous study \cite{zhu2014searching}. The HTRU dataset was obtained via private communication with C. Ng, while the FAST data was collected in this study.
Similar to Zhu et al. (2014) \cite{zhu2014searching}, we used the GBNCC dataset as an independent test dataset to verify the generalization of the models. 
Table \ref{tab:table0} displays the number of positive and negative examples in each dataset. In the GBNCC dataset, the positive candidates contained two types of labels, among which one indicated the fundamental signals of the pulsars, whereas the other indicated the harmonic signals of the pulsars.

\begin{tablehere}
\caption{Number of positive and negative examples in the datasets.}\label{tab:table0}
\vspace{-1mm}\footnotesize
\begin{center} \doublerulesep 0.1pt \tabcolsep 15pt
\begin{tabular}{lccccccc}
\hline
Dataset Names &  Positive &  Negative  &  Total \\
 \hline
PALFA & 3,951 & 6,672  & 10,623 \\
			HTRU & 903 & 271 & 1,174 \\
			FAST & 837 & 998 & 1,835 \\
			GBNCC & 277 & 89,731 & 90,008 \\

 \hline
\end{tabular}
\end{center}

\end{tablehere}

\begin{tablehere}
    \caption{Binary classification confusion matrix, which defines all the outcomes of predictions, including true negative (TN), false negative (FN), false positive (FP), and true positive (TP).}
	\label{tab:table1}
	\vspace{-1mm}\footnotesize
      \begin{center} \doublerulesep 0.1pt \tabcolsep 15pt
       \begin{tabular}{lccccc}
			\hline
			Outcomes & Negative Prediction & Positive Prediction \\
			\hline
			RFI & True Negative & False Positive \\
			True pulsar & False Negative & True Positive \\
			\hline
		\end{tabular}
	\end{center}
	
\end{tablehere}

We generally consider pulsar candidate selection to be a binary classification problem despite the fact that datasets occasionally contain more than two labels.
In our dataset, pulsars and their harmonic signals considered to be positive examples, while all the RFIs are considered to be negative examples. 
The evaluation metrics that are adopted for performing pulsar candidate classification are precision, recall and \textit{F$_1$} score. After defining a binary classification confusion matrix (Table \ref{tab:table1}), the metrics can be defined as follows:
\begin{equation}\rm{Precision} = \frac{\rm{TP}}{\rm{TP} + \rm{FP}},\end{equation}
\begin{equation}\rm{Recall} = \frac{\rm{TP}}{\rm{TP} + \rm{FN}},\end{equation}
\begin{equation}\textnormal{\rm{F$_1$ score}} = \frac{2 \times \rm{Precision} \times \rm{Recall}}{\rm{Precision} + \rm{Recall}}.\end{equation}

\subsection{Model training}

\begin{table*}
\caption{The number of training samples was continuously increased from 2,000 to 12,000. Five validation tests were performed for each data point, and the means of the \textit{F}$_1$ score were obtained.}\label{tab:table_curve}
\begin{center}\vspace{-2mm}\footnotesize \doublerulesep 0.2pt \tabcolsep 16pt
\begin{tabular*}{\textwidth}
{cccccccc}\toprule[0.65pt]  
&\multicolumn{6}{c} {Training dataset}
\\
\cline{2-7\ }
 &$N$=2000&$N$=4000& $N$=6000&
$N$=8000& $N$=10000& $N$=12000&
\\ \hline 
PICS Training \textit{F}$_1$ score&0.98&0.98&0.99&0.99&0.99&0.99\\
PICS--ResNet Training \textit{F}$_1$ score&0.98&0.98&0.98&0.98&0.98&0.98\\
PICS validation \textit{F}$_1$ score&0.86&0.90&0.91&0.91&0.92&0.92\\
PICS--ResNet validation \textit{F}$_1$ score&0.89&0.91&0.92&0.92&0.91&0.92\\
\bottomrule[0.65pt] 
\end{tabular*}
\end{center}
\end{table*}

In this subsection, we describe the training process of the 15-layer ResNet and PICS--ResNet models and observe the manner in which the models can converge. We used the same dataset, which included the PALFA, HTRU, and FAST dataset (Table \ref{tab:table0}) for training. There were a total of 13,632 labeled candidates in this dataset, among which 5,692 were pulsars (and their harmonic signals) and 7,940 were non-pulsars. 

\begin{figure}[H]
\centering
\includegraphics[scale=0.4]{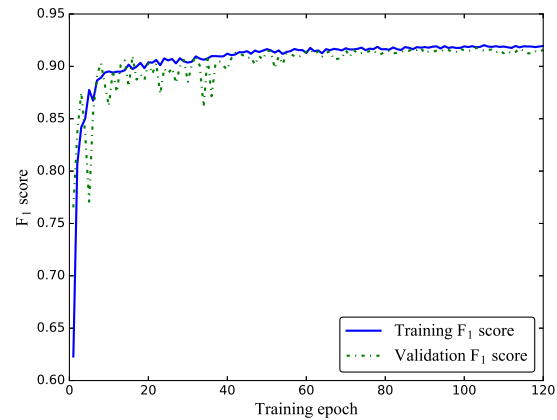}
\caption{Convergence of the 15 layers of ResNet over the training process. The x-axis represents the training epochs, whereas the y-axis represents the \textit{F}$_1$ score.}
\label{fig:resnet_learning}
\end{figure}

We used grid search to fine-tune the hyperparameters of the ResNet and selected the set of parameters leading to the optimal cross-validation \textit{F}$_1$ score and as well as a well-converging model. For the cross-validation tests, we randomly select 80\% of the entire dataset for training and the remaining 20\% for performing validation. Figure~\ref{fig:resnet_learning} illustrates the convergence of the 15 layers of the ResNet over the training process. The hyperparameters of the ResNet model were set to an initial learning rate of 0.01, a batch size of 32, a weight decay rate of 0.002, a ReLU leakiness of 0.1, and a training epoch of 80. 

\begin{figure}[H]
\centering
\includegraphics[scale=0.4]{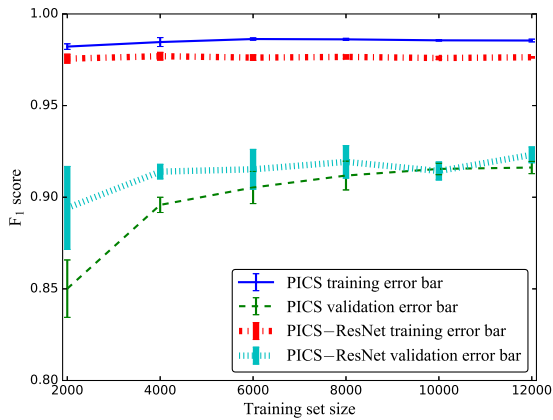}
\caption{The PICS--ResNet and PICS learning curves. The x-axis represents the training sample set from 2,000 to 12,000, whereas the y-axis represents \textit{F}$_1$ score.}
\label{fig:learning}
\end{figure}

The training and validation tests were performed to determine the manner in which the models converged and to identify the signs of overfitting. In the training process, we gradually increased the size of the training data and visualized the convergence of the training and validation \textit{F}$_1$ scores. We first used 2,000 samples to train the model, followed by the usage of 1,363 samples for validation. Next, we increased the training set size to 12,000. In each step, we randomly selected 1,363 samples as validation data; these data did not contain the training data. We performed five independent cross-validation tests to evaluate the reliability of the training process, with each test exhibiting a different random split of the training and validation data. The resulting \textit{F}$_1$ scores are presented in Table \ref{tab:table_curve} and the learning curves are plotted in Figure~\ref{fig:learning}. 

Figure~\ref{fig:learning} demonstrates that the training and validation scores of both the PICS and PICS--ResNet models improve as the size of the training set increases and that the PICS--ResNet model converges better than the PICS model.
The results indicate that the training size remains insufficient for the training and validation curve to completely converge and that some degree of overfitting still exists in the models. However, the PICS--ResNet demonstrates slightly superior performance and convergence when compared to those exhibited by the PICS.
The mean validation \textit{F}$_1$ score is 0.92 for PICS and 0.92 for PICS--ResNet.

\subsection{Generalization test using the GBNCC dataset \label{sec:gbncc}}

To independently validate our model, we applied it to a large set of GBNCC candidates that were excluded while training the model. Similar to the FAST, PALFA, and HTRU, the GBNCC survey used the \textsc{presto} search pipeline to generate candidates in the pfd file format; however, it used a low observing frequency ($\sim$300 MHz). Thus, the RFI environments are unique to the Green Bank telescope. All the candidates in the GBNCC dataset were verified and labeled by unprejudiced human experts; therefore, this dataset represents a realistic survey.  However, similar to any genuine survey candidate pool, the dataset is considerably imbalanced. There are 56 pulsars, 221 harmonics, and 89,731 RFI candidates (Table \ref{tab:table0}).  The result of the GBNCC dataset serve to measure the performance of our model in a realistic pulsar survey and to test its generalizability with respect to a different telescope, different observing frequency, and different RFI environment.

\begin{figure}[H]
\centering
\includegraphics[scale=0.4]{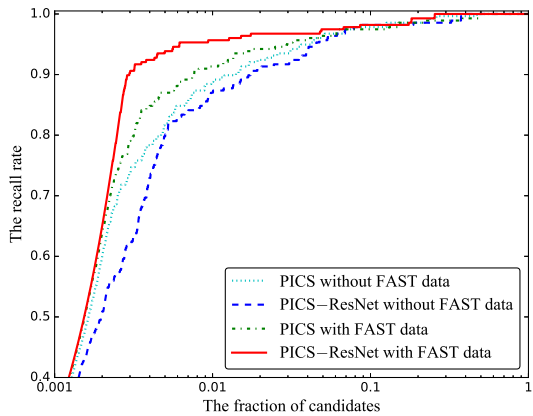}
\caption{PICS--ResNet and PICS recall curves. The x-axis represents the fraction of the candidates examined, whereas the y-axis represents the recall rate of the examined candidates. Here, the recall rate is calculated based on the fraction of pulsar signals (including 56 fundamental signals and 221 harmonic signals) ranked in the top $X$ fraction of the candidates.}
\label{fig:recall_curve}
\end{figure}

To determine whether the inclusion of new data improves the performance and generalization of the models, we conducted experiments using the GBNCC test data using the following two steps: 1. we trained our models using an old dataset that excluded the FAST data, and 2. we further trained the models using a large dataset that included FAST data. In training the PICS--ResNet model, we collected 11,797 training samples from the PALFA and the HTRU South survey (Table \ref{tab:table0}). The model ran for 80 epochs, usually with an iterative process and a checkpoint being saved after each epoch. In addition, we used a standard momentum optimizer as the learning algorithm for training in which the initial learning rate was set to 0.01, the weight decay was set to 0.002, and the batch size set to 32. 
We used the same training set to train the PICS model; the results are presented in Table \ref{tab:table2}. PICS sorted 52 pulsar fundamental signals and 194 harmonics into the top $1\%$ of 90,008 candidates, while PICS--ResNet sorted 51 fundamental signals and 190 harmonics into the top $1\%$. The recall curves of the two models are denoted in Figure~\ref{fig:recall_curve}. The total number of pulsar fundamental signals is 56, and the number of harmonics is 221. 
Table \ref{tab:table2} reveals that when trained with the old dataset, both PICS and PICS--ResNet reach a $\sim$90\% recall rate for the GBNCC test by considering only the top 1\% of the candidates.

\begin{tablehere}
    \caption{Classification results of the PICS and PICS--ResNet models using the GBNCC dataset when the models were trained with the PALFA and HTRU data but not the FAST data. Only the candidates who were ranked in the top $1\%$ of the dataset were selected, and the recall rates of the pulsars and their harmonics were calculated. }
	\label{tab:table2}
    \vspace{-1mm}\footnotesize
      \begin{center} \doublerulesep 0.1pt \tabcolsep 2pt
       \begin{tabular}{lccccccc}
		\hline
			Method & Fundamental (recall) & Harmonic (recall) & PSR (recall) \\
			\hline
			PICS (Top $1\%$) & 52 ($93\%$) & 194 ($88\%$) & 246 ($89\%$)\\
			PICS--ResNet (Top $1\%$) & 51 ($91\%$) & 190 ($86\%$) & 241 ($87\%$)\\
			\hline
		\end{tabular}
	\end{center}
	
\end{tablehere}

To dtermine the influence of FAST data on the performance of the models, we trained the models using additional data collected from the FAST drift-scan pulsar survey. 
A total of 13,632 samples (see Table \ref{tab:table0}) were used from the PALFA, HTRU, and FAST. The test results are presented in Table \ref{tab:table3}.
PICS--Resnet sorted 54 fundamental pulsars and 211 harmonics. Among the results, the top 210 candidates were all pulsars, with the first non-pulsar being the 211$^{th}$ candidate. There were 254 pulsars among the top 290 candidates. We used the same training data to train the PICS, and the resulting model sorted 52 fundamental pulsars and 201 of their harmonics into the top 1\% of all the candidates.
Table \ref{tab:table3} demonstrate that when trained with the FAST data, both PICS and PICS--ResNet exhibited superior performance, and PICS--ResNet reached a $\sim$96\% recall rate in the GBNCC test by considering only the top 1\% of all the candidates. Figure~\ref{fig:recall_curve} displays the recall curves for the two models trained using FAST data.
The results reveal that the inclusion of new data obtained from the FAST significantly improved the models' recall rates with respect to the harmonic signals of pulsars. We conclude that increasing the size and diversity of the training data can increase the generalization of the models.

\begin{tablehere}
\caption{Classification results of the PICS and PICS--ResNet models using the GBNCC dataset. The models were trained with datasets from the PALFA, HTRU, and FAST. Only the candidates ranked in the top $1\%$ of the dataset were selected, and the recall rates of the pulsars and their harmonics were calculated. }
	\label{tab:table3}
	\vspace{-1mm}\footnotesize
      \begin{center} \doublerulesep 0.1pt \tabcolsep 2pt
       \begin{tabular}{lccccccc}
		\hline
			Method & Fundamental (recall) & Harmonic (recall) & PSR (recall) \\
			\hline
			PICS (Top $1\%$) & 52 ($93\%$) & 201 ($91\%$) & 253 ($91\%$)\\
			PICS--ResNet (Top $1\%$) & 54 ($96\%$) & 211 ($96\%$) & 265 ($96\%$)\\
			\hline
		\end{tabular}
	\end{center}
	
\end{tablehere}

While testing using the GBNCC dataset, we also evaluated the runtime of the models.
It took the ResNet model $\sim$0.22 ms to process one 2D plot using a cluster of 24 2.7--GHz CPUs and two 1080Ti GPUs. In contrast, it took the CNN model of the PICS system $\sim$7.2 ms to process one 2D plot. It should be noted that our method ran efficiently because the CPU and GPU were running concurrently. It took PICS--ResNet 61 s and PICS 95 s to predict 10,000 candidates. Overall, it took  $\sim$79 minutes for PICS--ResNet and $\sim$90 minutes for PICS to complete the classification of 90,008 candidates. The PICS--ResNet model exhibits a slightly faster performance than that exhibited by the PICS model.  

\subsection{Testing using the FAST data}

We used new data collected from the FAST drift--scan survey to further verify the robustness of our model. We collected 317,018 candidates from the survey; 15,542 were labeled by human experts, among which 1,163 were pulsars or their harmonics (hereafter simply referred to as pulsars). Both the PICS and PICS--ResNet models were trained with 13,632 samples (Table \ref{tab:table0}), including 1,835 samples obtained from the FAST data. An additional 326 pulsar samples and 13,321 RFI samples from the FAST data were reserved for testing. Table \ref{tab:table4} reveals that PICS--ResNet identified 320 pulsars and missed six, while PICS identified 310 pulsars and missed 16. This demonstrates that when the FAST data is used, PICS can achieve a recall rate of 95\%, whereas PICS--ResNet can achieve a recall rate of 98\%. In this experiment, the recall rate was calculated using a score threshold of 0.5. This was because the FAST test data was not as imbalanced as the GBNCC data, and there would have been considerably few candidates in the top 1\% to include all the true positive samples. Therefore, the recall numbers in Tables \ref{tab:table4} and \ref{tab:table3} should not be directly compared.

\begin{tablehere}
    \caption{Number of pulsars identified by each model when tested using the FAST data. The test set comprised a total of 326 pulsar candidates.}
	\label{tab:table4}
	\vspace{-1mm}\footnotesize
      \begin{center} \doublerulesep 0.1pt \tabcolsep 8pt
       \begin{tabular}{lccccccc}
       \hline
			Method & Recognition pulsar & Missing pulsar & Recall \\
			\hline
			PICS & 310 & 16 & $95\%$\\
			PICS--ResNet & 320 & 6 & $98\%$\\
			\hline
		\end{tabular}
	\end{center}
	
\end{tablehere}

\section{Discussion and conclusion}\label{sec:4}
In this study, we propose PICS--ResNet, which is a new ensemble model for pulsar candidate selection. This model inherits a two-layer structure from PICS \cite{zhu2014searching} and replaces the lower-layer CNN classifier with a ResNet-based model for performing 2D data classification. The PICS--ResNet model uses 15 layers of a deep neural network to identify 2D subplots and utilizes SVM to identify a 1D subplot. Our experimental results demonstrate that the PICS--ResNet model converges better than the PICS model during training and performs better than the PICS model when trained with the FAST data. In addition, when the FAST data is considered in training, the resulting models exhibit improved preformance in a test using the GBNCC data and a test using the FAST data. When trained with the FAST data and tested with the GBNCC data, PICS--ResNet sorts 96\% pulsars into the top 1\% of all the candidates. In addition, an experiment using FAST pulsar survey data demonstrates that PICS--ResNet can identify 98\% of pulsars with a score threshold of 0.5. 

Our experiments demonstrate that the ResNet model itself does not display significant signs of overfitting while classifying the 2D image features (Figure \ref{fig:resnet_learning}); however, the ensemble model appears to display slight overfitting, as illustrated by the learning curve (Figure \ref{fig:learning}). This remains a caveat because it is helpful in practice to exhibit a considerable separation in final scores while applying this model to data. In addition, our experimental results demonstrate that the inclusion of the data obtained from the FAST enhances the recall rate of the models with respect to the pulsar (harmonic) signals. Therefore, we can conclude that the training data exhibit insufficient diversity and that gathering additional data is an effective method for improving the pulsar candidate selection accuracy. Further, the simulated data can also be used as a supplement to the real data obtained from various surveys. 

The PICS--ResNet model was developed using the TensorFlow platform, and our experiments revealed that it can classify more than 1.6 million candidates per day using a dual GPU 24-core desktop computer (see Section \ref{sec:gbncc} for more details). The development code of the PICS--ResNet model can be found at https://github.com/dzuwhf/PICS-ResNet. In addition, our labeled FAST data has been made public on https://github.com/dzuwhf/FAST\_label\_data.

\vspace*{2mm} \Acknowledgements{\bahao The authors thank the referee's constructive comments and suggestions. The research work is supported by National Key R$\&$D Program of China No. 2017YFA0402600 and Natural Science Foundation of Shandong (No.ZR2015FL006). This work supported by the CAS International Partnership Program No.114A11KYSB20160008, the Strategic Priority Research Program of the Chinese Academy of Sciences Grant No. XDB23000000. This project is also supported by National Natural Science Foundation of China Grant No. 61472043, 11743002, 11873067, 11690024, 11725313, the Joint Research Fund in Astronomy (U1531242) under cooperative agreement between the NSFC and CAS and National Natural Science Foundation of China (grant No.11673005). WWZ is supported by the Chinese Academy of Science Pioneer Hundred Talents Program. The authors also thank Chavonne Bowen and Alan Ho for labeling FAST pulsar candidates, the PALFA, GBNCC team and the Arecibo Remote Command Center (ARCC) students, Cherry Ng, Meng Yu, et al. for labeling and sharing their data. }

\end{multicols}
\end{document}